\documentstyle[12pt,epsf]{article}
\renewcommand{\baselinestretch}{1.2}
\newcommand{\uno}{{ 1\:\!\!\!\mbox{I}}}

\def\npb#1#2#3{    {\it Nucl. Phys. }{\bf B #1} (19#2) #3}
\def\plb#1#2#3{    {\it Phys. Lett. }{\bf B #1} (19#2) #3}
\def\prd#1#2#3{    {\it Phys. Rev. }{\bf D #1} (19#2) #3}
\def\prep#1#2#3{   {\it Phys. Rep. }{\bf #1} (19#2) #3}
\def\prl#1#2#3{    {\it Phys. Rev. Lett. }{\bf #1} (19#2) #3}

\def\ppnp#1#2#3{   {\it Prog. Part. Nucl. Phys. }{\bf #1} (19#2) #3}

\def\zpc#1#2#3{    {\it Zeit. f\"ur Physik }{\bf C #1} (19#2) #3}

\def\eq#1{{eq.~(\ref{#1})}}

\newcommand{\bea}{\begin{eqnarray}}
\newcommand{\beq}{\begin{equation}}
\newcommand{\eea}{\end{eqnarray}}
\newcommand{\eeq}{\end{equation}}

\newcommand{\K}{{\scriptscriptstyle K}}

\newcommand{\diag}{{\scriptscriptstyle{\rm{diag}}}}

\begin{document}
\pagestyle{empty}
\begin{flushright}
ROME1 -- 1109/95 \\
ROM2F/95/20 \\
September, 1995 \\
hep--ph/???
\end{flushright}
\centerline{\Large{\bf{Flavour changing neutral currents }}}
\centerline{\Large{\bf{and CP violating processes}} }
\centerline{\Large{\bf{in generalized supersymmetric theories}}}
\vskip 1cm
\centerline{\bf{ E. Gabrielli$^{a}$,
A. Masiero$^b$ and L. Silvestrini$^{c}$}}
\vskip .5cm
\centerline{$^a$ Dip. di Fisica,
Universit\`a di Roma ``La Sapienza" and}
\centerline{INFN, Sezione di Roma, P.le A. Moro 2, I-00185 Roma, Italy. }
\centerline{$^b$ Dip. di Fisica, Universit\`{a} di Perugia and}
\centerline{INFN, Sezione di Perugia, Via Pascoli, I-06100 Perugia, Italy.}
\centerline{$^c$ Dip. di Fisica, Univ. di Roma ``Tor Vergata''
and INFN, Sezione di Roma II,}
\centerline{Via della Ricerca Scientifica 1, I-00133 Roma, Italy.}
\begin{abstract}
We consider supersymmetric extensions of the standard model with general
non-universal soft breaking terms. We analyse in a model-independent way
the constraints on these terms at the electroweak energy scale coming from
gluino mediated
flavour (F) changing neutral current and CP-violating processes.
We have computed the complete $\Delta F=1$ and $\Delta F=2$  effective
hamiltonian for gluino mediated processes, including for the first time the
effect of box diagrams in the evaluation of $\epsilon^{\prime}/\epsilon$.
We present numerical results for the constraints on these non-universal soft
breaking terms for different values of the parameters,
extending the analysis also to the leptonic sector.
A comparison with previous results in the literature is given.
\end{abstract}
\vskip 3.0cm

\vfill\eject
\pagestyle{empty}\clearpage
\setcounter{page}{1}
\pagestyle{plain}
\newpage
\pagestyle{plain} \setcounter{page}{1}
Since the early advent of low energy supersymmetry (SUSY) \cite{susy1},
the flavour changing neutral current (FCNC) tests played a major role
in severely
constraining the SUSY mass spectrum, in particular the sfermion sector.
It is known that the major conclusion was the high degree of degeneracy which
is requested in the squark sector \cite{susy2,susyfcnc,susycp}.
As for CP violating aspects,
it turned out that the bound on the electric dipole moment of the
neutron prevents any large conspicuous effect due to genuinely SUSY
phases, while the SUSY CP violating contributions related to the usual CKM
phase are small in comparison to the standard model (SM) contribution at
least in the minimal SUSY model (MSSM) \cite{susycp}.
Nowadays there is a renewed interest in both the FCNC and CP
issues in SUSY.
This is mainly due to some progress that was made in the issue of SUSY
breakdown in effective supergravities which emerge as the low energy
limit of superstring theories and, consequently, additional information
on the soft SUSY breaking terms was gained \cite{string}.
It turns out that, somewhat worryingly, the generic pattern of such soft
breaking terms does not correspond to the usual universality
conditions which are at the basis of the high degeneracy in the
sfermion sector. Also the imaginary parts of these soft breaking
terms have no a priori reason to be so small.

To test those effective supergravities in their FCNC and CP violation
implications one needs a model independent parameterization of  FCNC.
A particularly interesting class of these contributions is provided by the
gluino-induced FCNC effects \cite{susyfcnc,susycp}.
The most efficient parameterization is obtained
in the so-called superKM basis where gluino-quark-squark couplings
are diagonal in flavour and all the flavour changing (FC) effects
are due to the non-diagonality of the squark mass matrices \cite{hall}.
As long as the ratio of the off-diagonal
entries over an average squark mass remains a small parameter, the
first term of the expansion which is obtained by an off-diagonal mass
insertion in the squark propagators represents a suitable approximation.
The method avoids the specific knowledge of the sfermion mass matrices.

In this letter we present the main results of a new analysis of the full SUSY
contributions due to gluino or neutralino exchange to the set of FCNC and
CP violating phenomena. The motivation for this study is twofold :
\begin{enumerate}
\item the two previous long analyses \cite{gabbmas,hagelin} of this kind,
whose results are currently
used in the literature, differ quantitatively in several points and
a definitive clarification of these controversial aspects is needed;
\item even more important, the constraints coming from CP violation
have so far been only partially included in the analysis.
\end{enumerate}
To be sure, CP violating phenomena were not considered at all in ref.
\cite{gabbmas},
while in ref. \cite{hagelin} the analysis of CP violation in the
$\Delta S=1$ sector
takes into account only the superpenguin contributions disregarding SUSY box
diagrams.
On the contrary, in our analysis we show that SUSY box contributions
to $\Delta S=1$ CP violation are of the same order as the superpenguins ones
and, indeed, the interference of these two classes of contributions leads to
results which may differ by one order of magnitude with respect to what
was previously found.

We briefly state the ground for our discussion.
It has been known for more than ten years now that gluino-quark-squark
$(\tilde{g}-q-\tilde{q})$ vertices can exhibit flavour change
\cite{susyfcnc,susycp}.
The point is
that in general the $q$ and $\tilde{q}$ mass matrices are not
simultaneously diagonalizable. This might be due to the
initial conditions: SUSY breaking terms may yield contributions which are
not universal \cite{string}, i.e. they are not proportional to the unit matrix
in
flavour space. Otherwise, even starting with universal mass contribution
to sfermions in the SUSY soft breaking sector, renormalization effects from
the starting point, i.e. the scale of supergravity breaking,
down to the Fermi scale
can bring about a misalignment between $q$ and $\tilde{q}$ mass matrices
\cite{susyfcnc,susycp}.
This latter situation is what we encounter in the minimal SUSY standard model.
For instance, consider the mass matrix squared
of the scalar partner of the left-handed down-quarks $d_L$.
At the scale of supergravity
breaking this matrix consists of the SUSY conserving contribution
$m_dm_d^{\dagger}$ (where $m_d$ denotes the down
quark mass matrix) and the SUSY
breaking universal contribution $\tilde{m}^2{\uno}$.
However, the term $h_uQHu^c$ of the superpotential
generates a logarithmically
divergent contribution which is proportional to
$h_uh_u^{\dagger}$ and, hence,
to $m_um_u^{\dagger}$ ($m_u$ being the up-quark mass matrix).
Hence the resulting $\tilde{d}_L$ mass matrix squared at the Fermi scale is:
\beq
m^2_{\tilde{d}_L\tilde{d}_L}=m_dm^{\dagger}_d+
\tilde{m}^2 {\uno} +c\,m_um^{\dagger}_u
\label{msdown}
\eeq
Clearly the $(\tilde{g}-d_L-\tilde{d}_L)$ couplings are no longer flavour
diagonal
given the presence of the last term in the r.h.s. of \eq{msdown}. The above
mentioned mass-insertion approximation results from choosing the
so-called super-KM matrix where one takes a basis for the $\tilde{d}_L$
so that the $\tilde{g}-d_L-\tilde{d}_L$ are flavour diagonal and the
FCNC effects are accounted for by flavour changing mass insertions
in the $\tilde{q}$ propagators. From \eq{msdown} it is easy to realize that the
mass insertion needed to accomplish the transition from $\tilde{d}_{iL}$
to $\tilde{d}_{jL}$ ($i,j$ flavour indices) is given by:
\beq
\left(\Delta_{LL}^d\right)_{ij}=
c\left[K\left(m_u^{\diag}
\right)^2K^{\dagger}\right]_{ij}
\label{massins}
\eeq
where $K$ is the Cabibbo-Kobayashi-Maskawa matrix and
$m_u^{\diag}$
denotes the diagonalized up-quark mass matrix.
In the following, with the notation
$\left(\Delta_{AB}^q\right)_{ij}$, we mean the mass insertion needed
for a transition from a squark $\tilde{q}_{iA}$ to $\tilde{q}_{jB}$
with $A=(L,R)$ and $B=(L,R)$.
Actually, as particularly
emphasized in ref. \cite{hagelin}, the above expression for
$(\Delta_{LL}^{d})_{ij}$
may be somewhat misleading since one might think that the term $c$
in the r.h.s. of \eq{massins} is a constant
(i.e. independent of the SUSY breaking scale) or at most depends
logarithmically on it. On the contrary, the one-loop RGE's show that $c$
depends
quadratically on that scale. Since also the average squark mass is proportional
to this scale, the meaningful parameter for our mass insertion approximation is
the dimensionless quantity $\delta=\Delta/\tilde{m}^2$,
where $\tilde{m}$ denotes at the same time the average squark mass
and the typical SUSY breaking scale. This observation is of utmost
relevance if one wants to understand the scaling of the SUSY contribution
to FCNC with increasing squark masses. The powers of $\tilde{m}$ in the
denominator which are present to compensate for $\Delta$ mass insertions
in the numerator do not have to be considered if also $\Delta$ is proportional
to $\tilde{m}^2$. This justifies why gluino-induced FCNC SUSY contributions
remain still sizeable even for $\tilde{q}$ masses above 1 TeV \cite{hagelin},
as we will see in what follows.

Two comments are in order before giving our results.
There are three classes of sfermion mixings according to the helicity of their
fermionic partners: $\Delta_{LL},~\Delta_{RR},~\Delta_{LR}$.
In the MSSM there is a sharp hierarchy among them \cite{gabbmas}.
Only $\Delta_{LL}$ appears as a simple mass insertion since the FC effect is
related to the property of the $\tilde{q}_L$ to sit in SU(2) doublets,
$\Delta_{LR}$ results from a FC $\Delta_{LL}$ insertion followed by a
$\Delta_{RR}$ flavour conserving insertion. Finally a $\Delta_{RR}$
FC contribution would require three mass insertions since only $\Delta_{LL}$
can yield a FC.
Hence $(\Delta_{LL})_{ij}>>(\Delta_{LR})_{ij}>>(\Delta_{RR})_{ij}$ with
$i \neq j$ in the MSSM. However, it should be clear from the above
sketched argument of the source of FC in the $\tilde{q}$ propagators in
the MSSM, that this conclusion is quite model-dependent.
In particular, if FC effects are produced by ``initial" conditions
one cannot make any general statement on the relative size of the
three contributions $\Delta_{LL},\Delta_{LR},\Delta_{RR}$.

The second observation is related to the appearance of the CKM elements
in the expressions of the $\Delta_{LL}$ in the MSSM (\eq{massins}).
In the gluino-induced FC contributions one obtains a GIM suppression
mechanism which is entirely analogue of what occurred in W-mediated
FCNC effects in the SM, in particular with the same CKM angles and phase.
This is crucial in understanding
the smallness of the SUSY contributions both in the FCNC and CP violating
phenomena. In particular, the smallness of the angles connecting transitions
between first and third generation is at the basis of the smallness also
of the gluino-induced SUSY contributions to CP violation in the Kaon
system (we are assuming here that the only source of CP violation in the MSSM
is
the CKM phase). Also this property is very specific of the MSSM structure,
in particular of the universality of its soft breaking terms (see the second
term in the r.h.s. of \eq{msdown}).

We now come to the results of our analysis concerning the terms
$(\Delta_{LL})_{ij}$, $(\Delta_{LR})_{ij}$ and $(\Delta_{RR})_{ij}$
in the u- and d-sectors. In the following we consider the case in which
$(\Delta_{LR})_{ij}\simeq (\Delta_{RL})_{ij}$.
We will comment later on the analogous contributions
in the charged lepton sector.

 \begin{table}
 \begin{center}
 \begin{tabular}{||c|c|c|c||}  \hline \hline
 $x$ & $\sqrt{\left|{\mbox Re}  \left(\delta^{d}_{12} \right)_{LL}^{2}\right|}
$
 &
 $\sqrt{\left|{\mbox Re}  \left(\delta^{d}_{12} \right)_{LR}^{2}\right|} $ &
 $\sqrt{\left|{\mbox Re}  \left(\delta^{d}_{12}
\right)_{LL}\left(\delta^{d}_{12}
 \right)_{RR}\right|} $ \\
 \hline
 $
   0.3
 $ &
 $
1.9\times 10^{-2}
 $ & $
7.9\times 10^{-3}
 $ & $
2.5\times 10^{-3}
 $ \\
 $
   1.0
 $ &
 $
4.0\times 10^{-2}
 $ & $
4.4\times 10^{-3}
 $ & $
2.8\times 10^{-3}
 $ \\
 $
   4.0
 $ &
 $
9.3\times 10^{-2}
 $ & $
5.3\times 10^{-3}
 $ & $
4.0\times 10^{-3}
 $ \\ \hline \hline
 $x$ & $\sqrt{\left|{\mbox Re}  \left(\delta^{d}_{13} \right)_{LL}^{2}\right|}
$
 &
 $\sqrt{\left|{\mbox Re}  \left(\delta^{d}_{13} \right)_{LR}^{2}\right|} $ &
 $\sqrt{\left|{\mbox Re}  \left(\delta^{d}_{13}
\right)_{LL}\left(\delta^{d}_{13}
 \right)_{RR}\right|} $ \\
 \hline
 $
   0.3
 $ &
 $
4.6\times 10^{-2}
 $ & $
5.6\times 10^{-2}
 $ & $
1.6\times 10^{-2}
 $ \\
 $
   1.0
 $ &
 $
9.8\times 10^{-2}
 $ & $
3.3\times 10^{-2}
 $ & $
1.8\times 10^{-2}
 $ \\
 $
   4.0
 $ &
 $
2.3\times 10^{-1}
 $ & $
3.6\times 10^{-2}
 $ & $
2.5\times 10^{-2}
 $ \\ \hline \hline
 $x$ & $\sqrt{\left|{\mbox Re}  \left(\delta^{u}_{12} \right)_{LL}^{2}\right|}
$
 &
 $\sqrt{\left|{\mbox Re}  \left(\delta^{u}_{12} \right)_{LR}^{2}\right|} $ &
 $\sqrt{\left|{\mbox Re}  \left(\delta^{u}_{12}
\right)_{LL}\left(\delta^{u}_{12}
 \right)_{RR}\right|} $ \\
 \hline
 $
   0.3
 $ &
 $
4.7\times 10^{-2}
 $ & $
6.3\times 10^{-2}
 $ & $
1.6\times 10^{-2}
 $ \\
 $
   1.0
 $ &
 $
1.0\times 10^{-1}
 $ & $
3.1\times 10^{-2}
 $ & $
1.7\times 10^{-2}
 $ \\
 $
   4.0
 $ &
 $
2.4\times 10^{-1}
 $ & $
3.5\times 10^{-2}
 $ & $
2.5\times 10^{-2}
 $ \\ \hline \hline
 \end{tabular}
 \caption[]{Limits on $\mbox{Re}\left(\delta_{ij}\right)_{AB}\left(
\delta_{ij}\right)_{CD}$, with $A,B,C,D=(L,R)$, for a squark mass
 $\tilde{m}=500\mbox{GeV}$ and for different values of
 $x=m_{\tilde{g}}^2/\tilde{m}^2$.}
 \label{kkba}
 \end{center}
 \end{table}

 \begin{table}
 \begin{center}
 \begin{tabular}{||c|c|c||}  \hline \hline
  & & \\
 $x$ & $\left|\left(\delta^{d}_{23} \right)_{LL}\right| $ &
 $\left|  \left(\delta^{d}_{23} \right)_{LR}\right| $ \\
  & & \\ \hline
 $
   0.3
 $ &
 $
4.4
 $ & $
1.3\times 10^{-2}
 $ \\
 $
   1.0
 $ &
 $
8.2
 $ & $
1.6\times 10^{-2}
 $ \\
 $
   4.0
 $ &
 $
26
 $ & $
3.0\times 10^{-2}
 $ \\ \hline \hline
 \end{tabular}
 \caption[]{Limits on the $\left| \delta_{23}^{d}\right|$ from
 $b\rightarrow s \gamma$ decay for a squark mass $\tilde{m}=500\mbox{GeV}$
 and for different values of $x=m_{\tilde{g}}^2/\tilde{m}^2$.}
 \label{bsg}
 \end{center}
 \end{table}

First we consider the CP conserving FCNC processes. In the down sector the
$\Delta_{ij}$ mass insertions are bounded by the $K-\bar{K}$
mass difference
$(\delta_{12})$, the $B_d-\bar{B}_d$ mixing $(\delta_{13})$ and
the branching ratio BR $(b\rightarrow s+\gamma)$ $(\delta_{23})$,
while the only available bound in the up-sector concerns
$\Delta_{12}$ from $D-\bar{D}$ mixing. We report our
results in tables \ref{kkba} and \ref{bsg}
for an average $\tilde{q}$ mass of 500 GeV and for different values of
$x=m^2_{\tilde{g}}/\tilde{m}^2$, where $m_{\tilde{g}}$ is the gluino mass.
Table \ref{bsg} shows that the decay
$(b\rightarrow s+\gamma)$ does not limit the $\delta_{LL}$ insertion
for a SUSY breaking of O(500 GeV). Indeed, even taking
$m_{\tilde{q}}=100\mbox{GeV}$, the term $(\delta_{23})_{LL}$ is only marginally
limited ( $(\Delta_{LL})_{23}<0.3$ for $x=1$).
Obviously, $(\delta_{23}^d)_{LR}$
is much more constrained since with a $\Delta_{LR}$ FC mass insertion the
helicity flip needed for $(b\rightarrow s+\gamma)$ is realized in the gluino
internal line and so this contribution has an amplitude enhancement
of a factor $m_{\tilde{g}}/m_b$ over the previous case with
$\Delta_{LL}$.

Concerning the calculation of the bounds in tables \ref{kkba} and \ref{bsg}
we find some discrepancies with previous results quoted in the literature.
In eqs. (3.2 a and c) of ref. \cite{gabbmas} the terms
proportional to
the function $M(x)$ must be multiplied by the coefficient ($-1/2$),
while in eq.
(3.2 b) the function $G(x)$ must be multiplied by ($-1$).
In eq. (4.2) of ref. \cite{hagelin}, the terms proportional
to $\Delta_{LL}\Delta_{RR} \cdot m_{\K}/(m_s+m_d)$ and
$\left(\Delta_{RL}\right)^2 \cdot m_{\K}/(m_s+m_d)$ must be multiplied by ($-
1$)
and ($-11/18$)
respectively, while the major difference concerns the contribution to
$\Delta m_{\K}$ proportional to $\Delta_{LR}\Delta_{RL}$ mass
insertion given by:
\beq
\Delta m_{\K}=\frac{\alpha_s^2}{216M^2_{\tilde{q}}}\frac{2}{3}
f_{\K}^2m_{\K}\frac{\delta\tilde{m}^2_{d_Ls_R}\delta\tilde{m}^2_{d_Rs_L}}
{M^4_{\tilde{q}}}
\tilde{f}_6(x)\left[84+144\left(\frac{m_{\K}}{m_s+m_d}\right)^2\right]
\eeq
where we have used the same notation of ref. \cite{hagelin} for comparison.
The complete expression for $\Delta m_K$ can be found in ref. \cite{proc},
together with more details on the analysis.

A similar analysis can be performed in the leptonic sector where the masses
$\tilde{m}$ and $m_{\tilde{g}}$ are replaced by the average slepton
mass and the photino mass $m_{\tilde{\gamma}}$ respectively.
A clear but important point to be stressed is that the
severe bounds that we provide
on the $\delta_{LL}$ and $\delta_{LR}$  mass insertions in the leptonic
sector and the consequent need for high degeneracy of charged sleptons,
only apply if separate lepton numbers are violated. It is well known
that in the MSSM the lepton numbers $L_{e},~L_{\mu}$ and $L_{\tau}$
are separately conserved because of the diagonality of the soft
breaking terms and the masslessness of neutrinos.
If at least one of this two properties is not present one can have
partial lepton number violation. A particularly interesting example is
the case where neutrinos acquire a mass through a see-saw mechanism
(for its SUSY version and the implications for FCNC see \cite{borzmas}).
In table \ref{lep} we exhibit the bounds on $\delta_{LL}^l$ and $\delta_{LR}^l$
coming from the limits on
$\mu\rightarrow e\gamma,~\tau\rightarrow e\gamma$ and
$\tau\rightarrow \mu\gamma$, for a slepton mass of O(100 GeV)
and for different values of $x=m_{\tilde{\gamma}}^2/\tilde{m}^2$.
Our results confirm those obtained in refs \cite{gabbmas,hagelin}.


 \begin{table}
 \begin{center}
 \begin{tabular}{||c|c|c||}  \hline \hline
  & & \\
 $x$ & $\left|\left(\delta^{l}_{12} \right)_{LL}\right| $ &
 $\left|  \left(\delta^{l}_{12} \right)_{LR}\right| $ \\
  & & \\ \hline
 $
   0.3
 $ &
 $
4.1\times 10^{-3}
 $ & $
1.4\times 10^{-6}
 $ \\
 $
   1.0
 $ &
 $
7.7\times 10^{-3}
 $ & $
1.7\times 10^{-6}
 $ \\
 $
   5.0
 $ &
 $
3.2\times 10^{-2}
 $ & $
3.8\times 10^{-6}
 $ \\ \hline \hline
  & & \\
 $x$ & $\left|\left(\delta^{l}_{13} \right)_{LL}\right| $ &
 $\left|  \left(\delta^{l}_{13} \right)_{LR}\right| $ \\
  & & \\ \hline
 $
   0.3
 $ &
 $
15
 $ & $
8.9\times 10^{-2}
 $ \\
 $
   1.0
 $ &
 $
29
 $ & $
1.1\times 10^{-1}
 $ \\
 $
   5.0
 $ &
 $
1.2\times 10^{2}
 $ & $
2.4\times 10^{-1}
 $ \\ \hline \hline
  & & \\
 $x$ & $\left|\left(\delta^{l}_{23} \right)_{LL}\right| $ &
 $\left|  \left(\delta^{l}_{23} \right)_{LR}\right| $ \\
  & & \\ \hline
 $
   0.3
 $ &
 $
2.8
 $ & $
1.7\times 10^{-2}
 $ \\
 $
   1.0
 $ &
 $
5.3
 $ & $
2.0\times 10^{-2}
 $ \\
 $
   5.0
 $ &
 $
22
 $ & $
4.4\times 10^{-2}
 $ \\ \hline \hline
 \end{tabular}
 \caption[]{Limits on the $\left| \delta_{ij}^{d}\right|$ from
 $l_j\rightarrow l_i \gamma$ lepton decay for
 a slepton mass $\tilde{m}=100\mbox{GeV}$ and for different values of
 $x=m_{\tilde{\gamma}}^2/\tilde{m}^2$.}

 \label{lep}
 \end{center}
 \end{table}

We tackle now the subject of one-loop CP violating contributions through
gluino exchange. As for the $\Delta S=2$ transitions, the corresponding
bounds on $\mbox{Im}(\delta_{12}^d)_{LL}^2$ and
$\mbox{Im}(\delta_{12}^d)_{LR}^2$ are readily obtained from those derived
for the real parts from $\Delta m_K$.
What is actually new in our analysis with respect to ref. \cite{hagelin},
but more generally, with respect to all previous works dealing with
gluino-induced
CP violation, is our treatment of the SUSY contributions to direct CP
violation in $\Delta S=1$ processes. Indeed only superpenguins were considered
to be relevant for $\epsilon^{\prime}$, while we obtain that also box diagrams
with gluinos exchange (see fig.1) give a sizeable contributions with
relevant interference effects with the superpenguins.

The contribution to the effective Hamiltonian for $\Delta S=1$ transitions
given by the gluino penguins and box diagrams can be written as
\begin{equation}
\cal{H}_{\scriptscriptstyle eff}=\sum_{i=3,7} \left\{
C_{\scriptscriptstyle i}O_{\scriptscriptstyle i}+
\tilde{C}_{\scriptscriptstyle i}\tilde{O}_{\scriptscriptstyle i}\right\}
\end{equation}
where we have chosen a basis of local four-fermion operators
$O_{\scriptscriptstyle i}$:
\begin{eqnarray}
O_{ 3} &=& ({\bar d}^{\alpha}_{L}\gamma^{\mu} s^{\alpha}_{L})
    \sum_{q=u,d,s}({\bar q}^{\beta}_{L}\gamma_{\mu}q^{\beta}_{L})
\nonumber \\
O_{ 4} &=& ({\bar d}^{\alpha}_{L}\gamma^{\mu} s^{\beta}_{L})
    \sum_{q=u,d,s}({\bar q}^{\beta}_{L}\gamma_{\mu}q^{\alpha}_{L})
\nonumber \\
O_{ 5} &=& ({\bar d}^{\alpha}_{L}\gamma^{\mu} s^{\alpha}_{L})
    \sum_{q=u,d,s}({\bar q}^{\beta}_{R}\gamma_{\mu}q^{\beta}_{R})
\nonumber \\
O_{ 6} &=& ({\bar d}^{\alpha}_{L}\gamma^{\mu} s^{\beta}_{L})
    \sum_{q=u,d,s}({\bar q}^{\beta}_{R}\gamma_{\mu}q^{\alpha}_{R})
\nonumber \\
O_{ 7} &=& \frac{g}{8\pi^2}{\bar d}^{\alpha}_{R}\sigma^{\mu\nu}
           t^A_{\alpha\beta}s^{\beta}_{L}G^A_{\mu\nu}
\label{basis}
\end{eqnarray}
where the operators $\tilde{O}_i$ can be obtained from $O_i$ by the exchange
$L \leftrightarrow R$.
Here $q_{R,L}=\frac{(1\pm\gamma_5)}{2}q$, $\sigma^{\mu\nu}=\frac{i}{2}
[\gamma^{\mu},\gamma^{\nu}]$, $\alpha$ and $\beta$ are colour
indices and $g$ is the strong coupling.
The colour matrices normalization is $ Tr(t^A t^B)= \delta^{AB}/2$.
\\
The Wilson coefficients are given by:
\begin{eqnarray}
C_3 &=& \frac{\alpha_s^2}{\tilde{m}^2}(\delta^d_{12})_{LL} \left(  -
\frac{1}{9}
{\mbox B}_1(x) - \frac{5}{9}{\mbox B}_2(x) - \frac{1}{18}{\mbox P}_1(x) -
\frac{1}{2}{\mbox P}_2(x)\right)
\nonumber  \\
C_4 &=& \frac{\alpha_s^2}{\tilde{m}^2}(\delta^d_{12})_{LL} \left(  -
\frac{7}{3}
{\mbox B}_1(x) + \frac{1}{3}{\mbox B}_2(x) + \frac{1}{6}{\mbox P}_1(x) +
\frac{3}{2}{\mbox P}_2(x)\right)
\nonumber  \\
C_5 &=& \frac{\alpha_s^2}{\tilde{m}^2}(\delta^d_{12})_{LL} \left(
\frac{10}{9}
{\mbox B}_1(x) + \frac{1}{18}{\mbox B}_2(x) - \frac{1}{18}{\mbox P}_1(x) -
\frac{1}{2}{\mbox P}_2(x)\right)
\nonumber  \\
C_6 &=& \frac{\alpha_s^2}{\tilde{m}^2}(\delta^d_{12})_{LL} \left(  -
\frac{2}{3}
{\mbox B}_1(x) + \frac{7}{6}{\mbox B}_2(x) + \frac{1}{6}{\mbox P}_1(x) +
\frac{3}{2}{\mbox P}_2(x)\right)
\nonumber  \\
C_7 &=& \frac{\alpha_s \pi}{\tilde{m}^2}\left[(\delta^d_{12})_{RR} \, \,
m_s \left(  - \frac{1}{3} {\mbox M}_3(x) - 3{\mbox M}_4(x) \right)\right.
\nonumber \\
&+&\left.(\delta^d_{12})_{RL} \,
\, m_{\tilde{g}} \left( - \frac{1}{3}{\mbox M}_1(x) -
3{\mbox M}_2(x)\right) \right]
\end{eqnarray}
where again the coefficients $\tilde{C}_i$ can be obtained from the $C_i$
just by the exchange $L \leftrightarrow R$, $x=m_{\tilde{g}}^2/
\tilde{m}^2$, and $m_s$ is the mass of the strange quark.
\\
The functions $B_i(x)$ which result from the calculation of the box diagrams
are given by:
\begin{eqnarray}
{\mbox B}_{1}(x) &=& \frac{1 + 4x - 5x^2 + 4x\ln (x) + 2x^2\ln (x)}
{8(1 - x)^4}
\nonumber \\
{\mbox B}_{2}(x) &=& x\: \frac{5 - 4x - x^2 + 2\ln(x) + 4x\ln(x)}
{2(1 - x)^4}
\label{bfunc}
\end{eqnarray}
while the functions $P_i(x)$ and $M_i(x)$ of the superpenguins can be derived
from ref. \cite{hagelin}.
In particular our results for the superpenguins coincide with those of ref.
\cite{hagelin}.

 \begin{table}
 \begin{center}
 \begin{tabular}{||c|c|c|c|c|c||}  \hline \hline
  $x$ & ${\scriptstyle\left|{\mbox Im} \left(\delta^{d}_{12}  \right)_{LL}
\right|} $ &
 ${\scriptstyle\left|{\mbox Im} \left(\delta^{d}_{12}\right)  _{LR}\right| }$ &
 ${\scriptstyle\sqrt{\left|{\mbox Im}  \left(\delta^{d}_{12} \right)_{LL}^{2}
\right|} }$ &
 ${\scriptstyle\sqrt{\left|{\mbox Im}  \left(\delta^{d}_{12} \right)_{LR}^{2}
\right|} }$ &
 ${\scriptstyle\sqrt{\left|{\mbox Im}  \left(\delta^{d}_{12}
\right)_{LL}\left(\delta^{d}_{12}
 \right)_{RR}\right|} }$ \\
 \hline
 $
   0.3
 $ &
 $
1.0\times 10^{-1}
 $ & $
1.1\times 10^{-5}
 $ &
 $
1.5\times 10^{-3}
 $ & $
6.3\times 10^{-4}
 $ & $
2.0\times 10^{-4}
 $ \\
 $
   1.0
 $ &
 $
4.8\times 10^{-1}
 $ & $
2.0\times 10^{-5}
 $ &
 $
3.2\times 10^{-3}
 $ & $
3.5\times 10^{-4}
 $ & $
2.2\times 10^{-4}
 $ \\
 $
   4.0
 $ &
 $
2.6\times 10^{-1}
 $ & $
6.3\times 10^{-5}
 $ &
 $
7.5\times 10^{-3}
 $ & $
4.2\times 10^{-4}
 $ & $
3.2\times 10^{-4}
 $ \\ \hline \hline
 \end{tabular}
 \caption[]{Limits on  $\mbox{Im}\left(\delta_{12}^{d}\right)_{AB}$ and on
$\mbox{Im}\left(\delta_{12}^{d}\right)_{AB}\left(\delta_{12}^{d}\right)_{CD}$,
with $A,B,C,D=(L,R)$, for
 a squark mass $\tilde{m}=500\mbox{GeV}$ and for different values of
 $x=m_{\tilde{g}}^2/\tilde{m}^2$.}

 \label{im}
 \end{center}
 \end{table}

In table \ref{im} we give the bounds on the imaginary parts of
$(\delta_{12}^d)_{LL},~(\delta_{12}^d)_{LR}$ (from $\Delta S=1$ transitions),
$(\delta_{12}^d)^2_{LL},~(\delta_{12}^d)^2_{LR}$ and
$(\delta_{12}^d)_{LL}(\delta_{12}^d)_{RR}$ (from $\Delta S=2$ transitions) for
an average squark mass $\tilde{m}=500\mbox{GeV}$.
In figs. 2 and 3 we exhibit the behaviour
of the upper bound of $\mbox{Im}(\delta_{12})_{LL}$ and
$\mbox{Im}(\delta_{12})_{LR}$  as a function of
$x=m_{\tilde{g}}^2/\tilde{m}^2$ and for $\tilde{m}=100 \mbox{GeV}$.
These bounds are obtained for an upper limit of
$\epsilon^{\prime}/\epsilon=10^{-3}$.
The effect of the interference between penguin and box contributions is
particularly severe for certain ranges of $x$. For instance, this is what
occurs
in the proximity of $x=1$ for the bounds in
$\mbox{Im}(\delta_{12})_{LL}$  (fig. 2).
The complete expression for the separate box and penguin contributions to
the $\Delta S=1$ effective hamiltonian will be given elsewhere \cite{ggms}.

{}From the results in table \ref{im} it is clear that, if one wishes to obtain
a sizeable contribution to $\epsilon^{\prime}$ from one-loop gluino exchange,
then $\mbox{Im}(\delta_{12})_{LL}$  should be of O($10^{-1}$). Moreover,
to respect the bound $\mbox{Im}(\delta_{12})^2_{LL}<10^{-6}$ from $\epsilon$,
$\mbox{Re}(\delta_{12})_{LL}$ should be extremely small.
Hence, unless $(\delta_{12})_{LL}$ is essentially imaginary
and taking into account the bound from $\epsilon$,
no large contribution to $\epsilon^{\prime}$ can arise from
$(\delta_{12})_{LL}$ mass insertion.
On the other hand, it is also apparent from table \ref{im} that a conspicuous
contribution to $\epsilon^{\prime}$, coming from $\mbox{Im}(\delta_{12})_{LR}$,
can arise in models with sizeable $\delta_{LR}$ mass insertions
without conflicting with the bound on this quantity coming from $\epsilon$.

Hence, although quantitatively the inclusion of the $\Delta S=1$ box
contributions changes the results of ref. \cite{hagelin}, we confirm the
main qualitative remark that those authors make, i.e. that SUSY models
with predominantly $\delta_{LL}$ or $\delta_{RR}$ contributions
to CP violation (such as the MSSM) are likely to be superweak,
while models with sizeable contribution to CP violation through
$\delta_{LR}$  mass insertions tend to be milliweak.
The implications of these considerations on SUSY contributions to CP violation
in models with non-universal soft breaking terms are presently under study
\cite{ggms}.
\section*{Acknowledgements}
We would like to thank G. Giudice for interesting discussions and the
physics department of
Padova for its kind hospitality. One of us,
 L.S.,  would like to acknowledge useful discussions
with R. Petronzio.
\renewcommand{\baselinestretch}{1}


\begin{figure}
\epsfysize=9.9cm
\epsfxsize=13.2cm

\epsffile{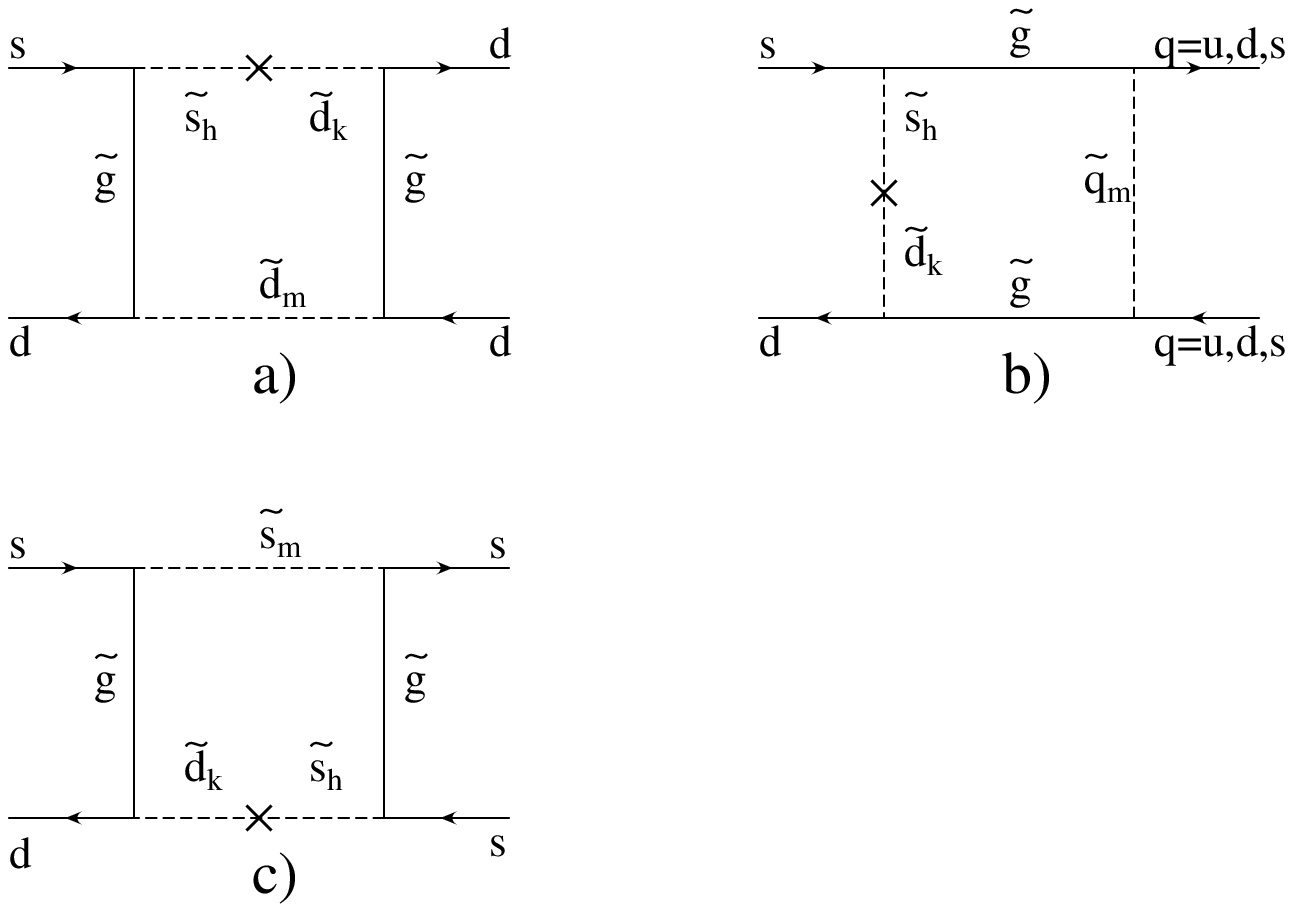}
\caption{The gluino box diagrams for $\Delta S=1$ transitions.}
\end{figure}

\newpage


\begin{figure}[t]   
    \begin{center}
    \epsfysize=12truecm
    \leavevmode\epsffile{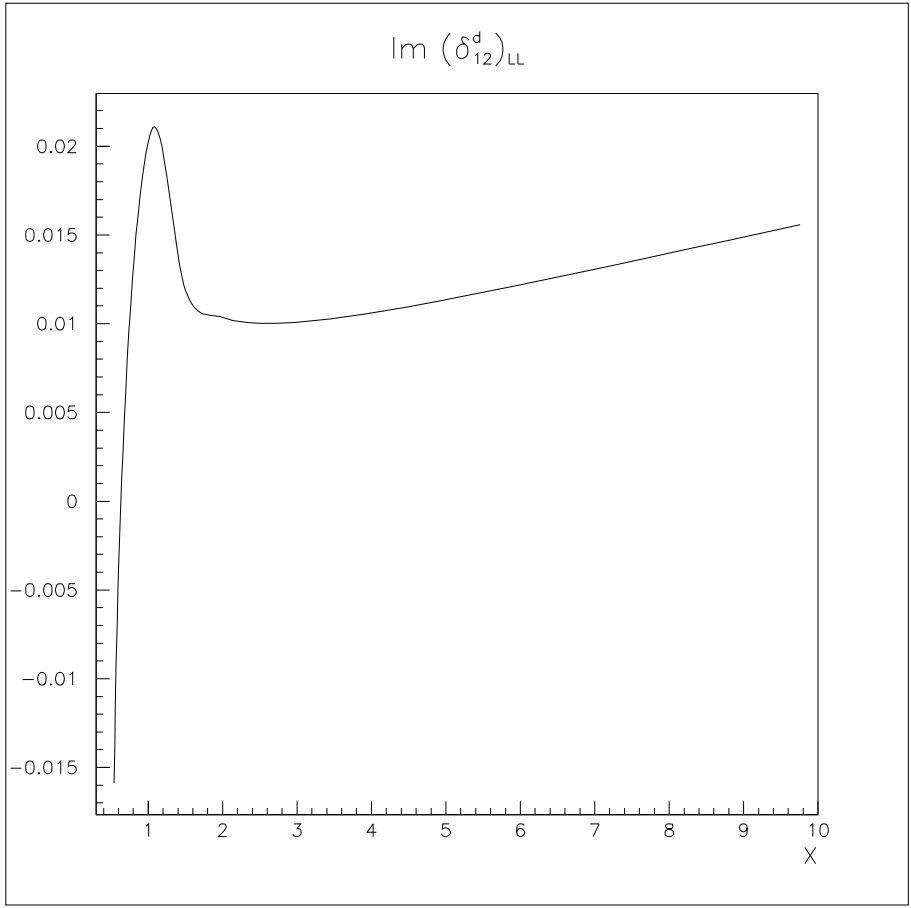}
    \end{center}
    \caption[]{The $\mbox{Im}\left(\delta^d_{12}\right)_{LL}$ as a function
     of $x=m_{\tilde{g}}^2/\tilde{m}^2$, for  a squark mass
     $\tilde{m}=100\mbox{GeV}$.}
\end{figure}
\newpage


\begin{figure}[t]   
    \begin{center}
    \epsfysize=12truecm
    \leavevmode\epsffile{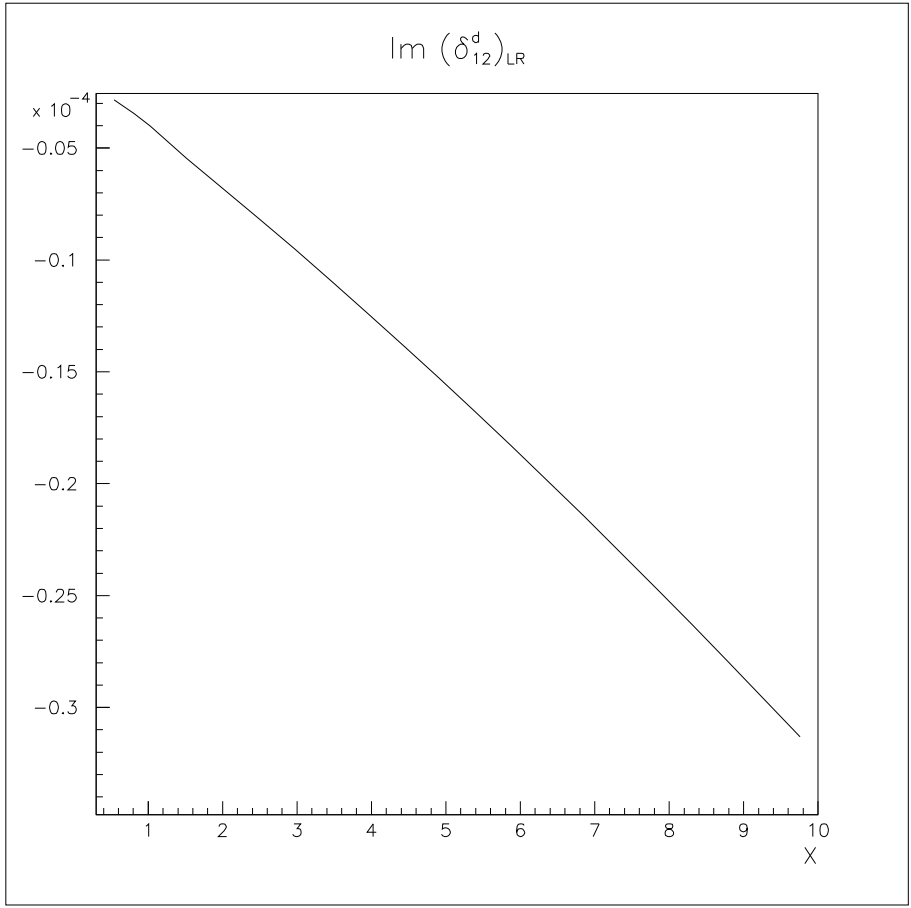}
    \end{center}
    \caption[]{The $\mbox{Im}\left(\delta^d_{12}\right)_{LR}$ as a function
     of $x=m_{\tilde{g}}^2/\tilde{m}^2$, for  a squark mass
     $\tilde{m}=100\mbox{GeV}$.}
\end{figure}
\end{document}